\documentclass[fleqn,12pt,twoside]{article}
\usepackage{espcrc1}

\usepackage{graphicx}

\title{Inclusive $\pi^{0}$ spectra at high transverse momentum in d-Au collisions at RHIC}

\author{A. Mischke\address{NIKHEF, Amsterdam, The Netherlands}\footnote{Present address:
Department of Subatomic Physics, University of Utrecht, The Netherlands.}
for the STAR Collaboration
\footnote{For the full author list and acknowledgments, see Ref.~\cite{SuppPartSpez04}.}}  

\begin{document}
\maketitle

\begin{abstract}

Preliminary results on inclusive neutral pion production in d-Au collisions at 
$\sqrt{s_{\rm NN}}$ = 200~GeV in the pseudo-rapidity range $0<\eta<1$ are 
presented. The measurement is performed using the STAR Barrel Electromagnetic 
Calorimeter (BEMC).
In this paper, the analysis of the first BEMC hadron measurement is described 
and the results are compared with earlier RHIC findings.
The $\pi^{0}$ invariant differential cross sections show good agreement 
with next-to-leading order (NLO) perturbative QCD calculations.

\end{abstract}

\section{Introduction}

A suppression of high $p_T$ hadron production relative to a simple scaling from p-p has been
observed in central Au-Au collisions at RHIC~\cite{SuppPartSpez04,SuppPart03}. 
It was also found that jet-like correlations opposite to trigger jets are suppressed and that 
the elliptic anisotropy in hadron emission persists out to very high 
$p_T$~\cite{BBCorr03,Flow03}.
In contrast, no suppression effects were seen in d-Au collisions, which provide an important 
control experiment for the effects of cold nuclear matter.
This has led to the conclusion that the observations are due to the high density final state 
in Au-Au. The most probable explanation to date is parton energy loss from induced gluon 
radiation (jet quenching). For a recent review see Ref.~\cite{Jac04}.
To quantitatively understand the existing modification from cold nuclear matter, precise
measurements of identified hadrons at high $p_T$ in d-Au are required. The STAR EMC allows
high transverse momentum measurements of $\pi^0$, $\eta$ mesons and direct photons and may 
contribute to the identification of $\rho$ mesons. 
In this paper we present preliminary results of neutral pion production in d-Au collisions.

\section{Experimental setup}

The results were obtained using two components of the STAR detector 
system~\cite{NimSTAR03}, namely the Time Projection Chamber (TPC) and the Barrel 
Electromagnetic Calorimeter (BEMC).
The TPC is situated in a 0.5 Tesla solenoidal magnetic field and provides a precise 
measurement of the charged particle trajectories.
The BEMC~\cite{NimEmc03} is a lead-scintillator sampling calorimeter with a of depth 
of 21 radiation lengths ($X_0$) and an inner radius of 220 cm, divided into towers 
of granularity $(\Delta\eta,\Delta\phi) = (0.05,0.05)$. 
Two layers of gaseous shower maximum detectors (SMD), located at a depth of 5 $X_0$, 
measure the EM shower shape with high resolution $(\Delta\eta,\Delta\phi) = (0.007,0.007)$. 
In this analysis a partial implementation of the BEMC was used, consisting of 2400 
towers covering $0<\eta<1$ and full azimuth (the deuteron beam has positive rapidity). 
When complete, the BEMC will have a coverage of $-1<\eta<1$.
Beam test results~\cite{NimAbsEnCal02} provide the absolute energy calibration, whereas 
the relative calibration is obtained from the peak position of minimum ionizing 
particles (mostly charged hadrons) on a tower-by-tower basis~\cite{EtSTAR04}. 
Moreover, an overall gain calibration was obtained by comparing the momentum of 
electrons identified in the TPC with the energy deposited in the BEMC.
The achieved energy resolution is $\delta E/E \approx 16\%/\sqrt{E}$.

In the minimum bias trigger a neutron signal was required in the Zero Degree 
Calorimeter (ZDC) in the Au beam direction resulting in an acceptance of 
$(95\pm3)\%$ of the d-Au hadronic cross section. 
To enhance the high $p_T$ range, two high tower triggers (HT1 and HT2) were used with 
an energy threshold of 2.5 GeV and 5 GeV for the highest EMC cluster energy, 
respectively. The tower occupancy is 1--5\% for d-Au events, and the high tower trigger 
efficiency is nearly 100\%.

\section{Data analysis}
\begin{figure}[t]
\begin{minipage}[b]{5cm}
 \begin{center}
 \includegraphics[width=5cm,height=6.2cm]{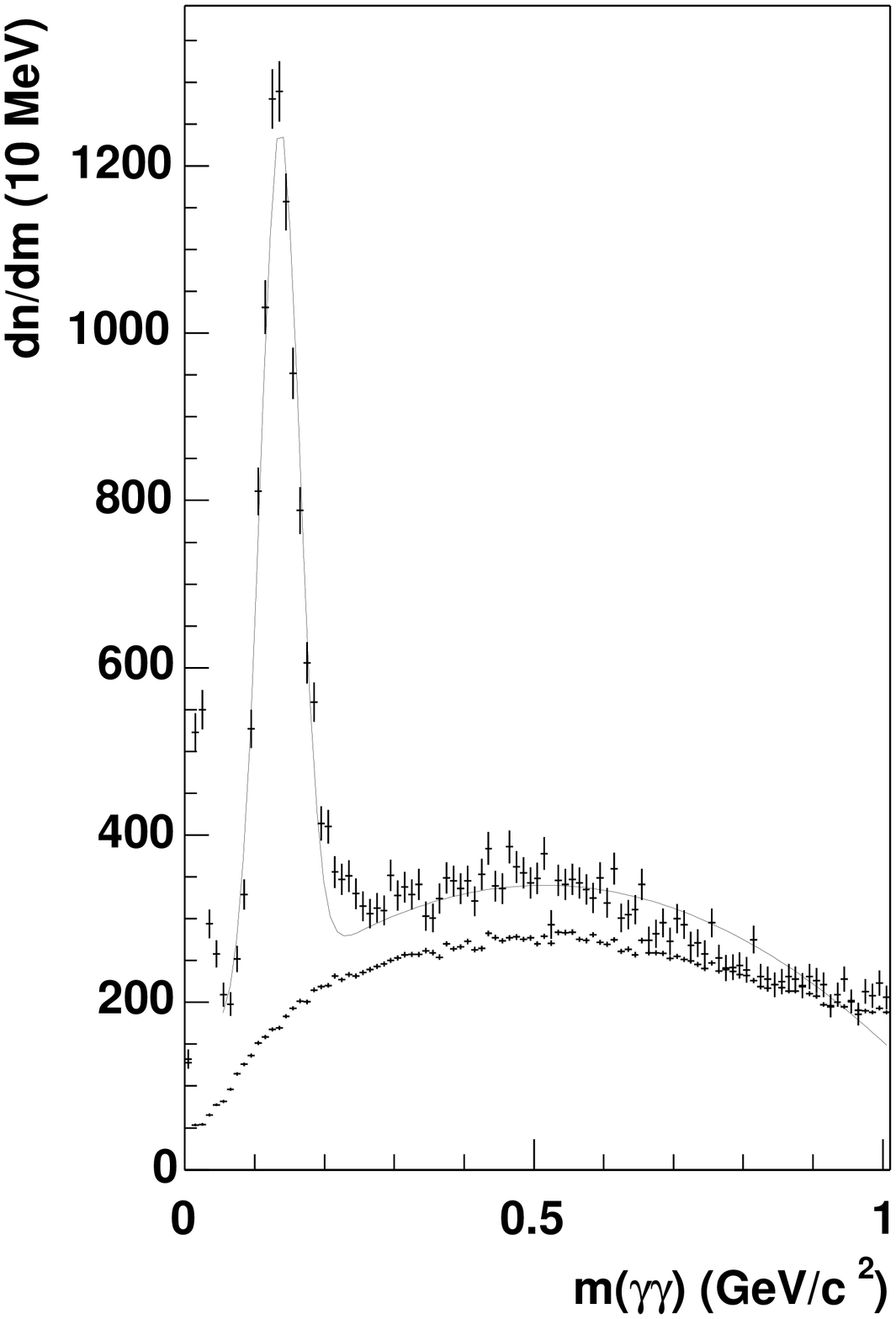}
 \end{center}
\end{minipage}
\hspace{\fill}
\begin{minipage}[b]{5cm}
 \begin{center}
 \includegraphics[width=5cm,height=6.2cm]{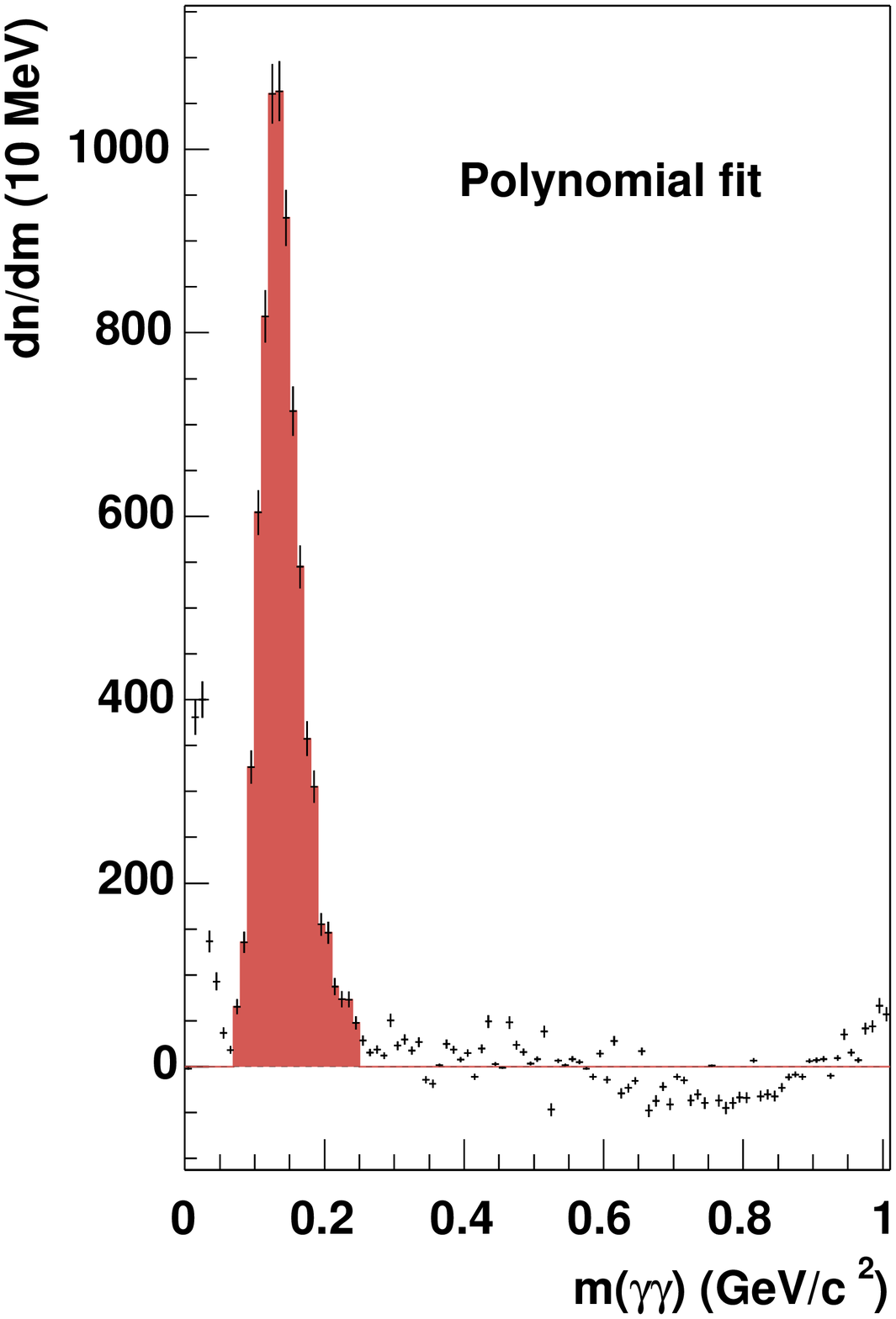}
 \end{center}
\end{minipage}
\hspace{\fill}
\begin{minipage}[b]{5cm}
 \begin{center}
 \includegraphics[width=5cm,height=6.2cm]{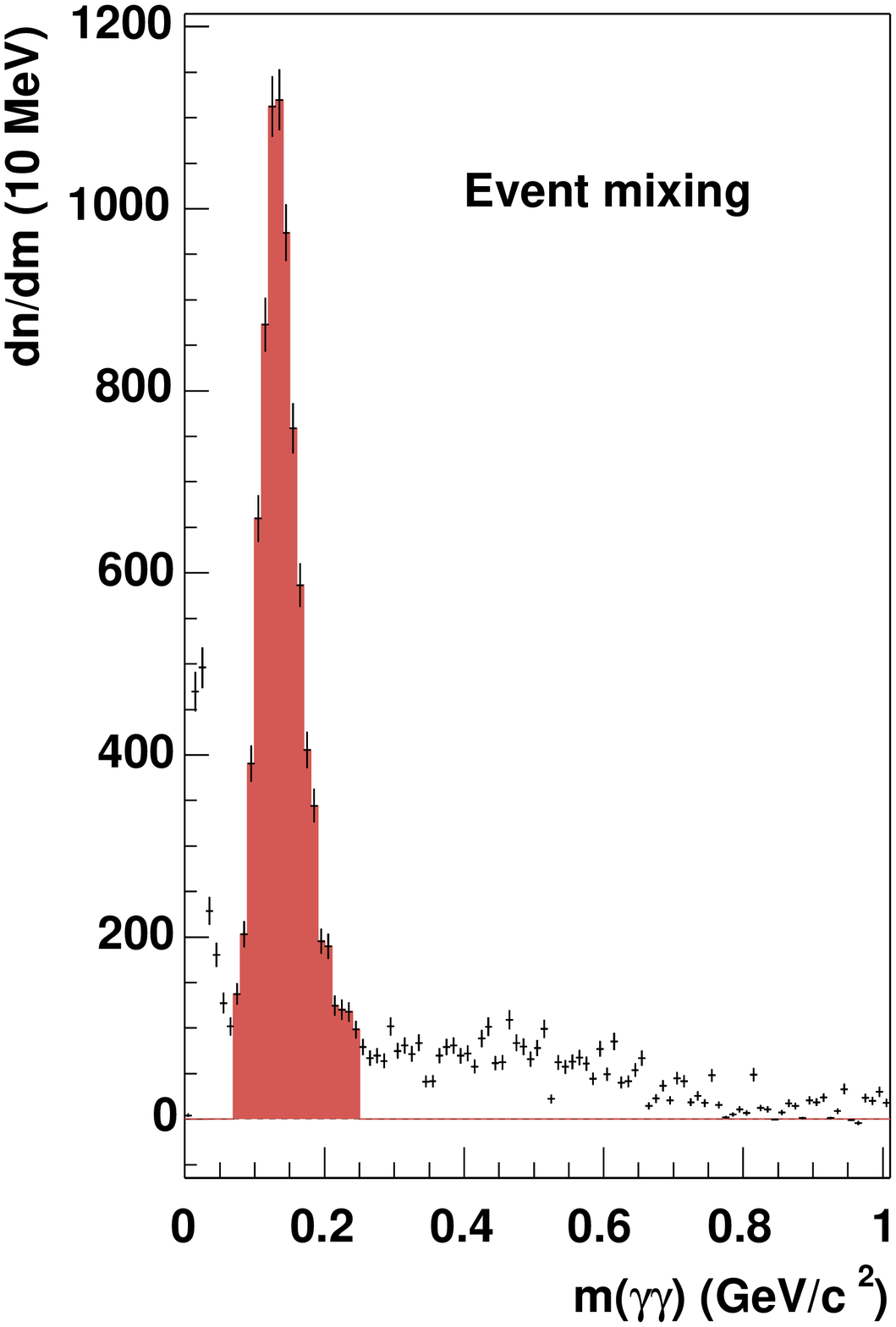}
 \end{center}
\end{minipage}
 \vspace{-1.cm}
 \caption{\protect \footnotesize 
Invariant mass distribution of neutral pion candidates in minimum bias d-Au 
reactions at $\sqrt{s_{\rm NN}}$ = 200~GeV (left). The solid line is a Gaussian 
plus second order polynomial fit and the lower histogram is obtained from 
event mixing.
The middle (right) figure is the background subtracted spectrum using the 
polynomial fit (event mixing). The relative mass resolution is 20\%.\vspace{-0.3cm}}
 \label{fig1}
\end{figure}
After event quality cuts, 10M d-Au events taken in the year 2003 were used for 
the analysis presented in this paper. 
Neutral pions were reconstructed in the decay channel $\pi^0\rightarrow\gamma\gamma$
(branching fraction 98.8\%) by calculating the invariant mass of all pairs of 
clusters in the calorimeter.
Only those clusters were selected which do not have a TPC track pointing to them. 
Furthermore, a cut on the energy asymmetry was imposed \mbox{$|E_1-E_2|/(E_1+E_2)\le0.5$}. 
At present, the full calibration of the calorimeter and the finding of noisy and 
dead towers are in progress.
To perform a neutral pion analysis at this stage a sub-sample of good towers was 
selected.
The quality of the individual towers was checked using the $\pi^0$ invariant mass 
distribution. The towers are assigned by the decay photons with the highest energy.
Only those towers were used which have a well defined $\pi^0$ signal above 
the combinatorial background. 
A Gaussian fit to the $\pi^0$ signal has to have a relative mass and width 
error of 30\% and 50\%, respectively.
By this method approximately one third of all the towers was used for the present
analysis. 
The $\pi^0$ peak position of all good towers was used to perform an additional 
tower gain correction (7\% on average).
\begin{figure}[t]
\begin{minipage}[b]{6cm}
 \begin{center}
 \includegraphics[height=8.8cm]{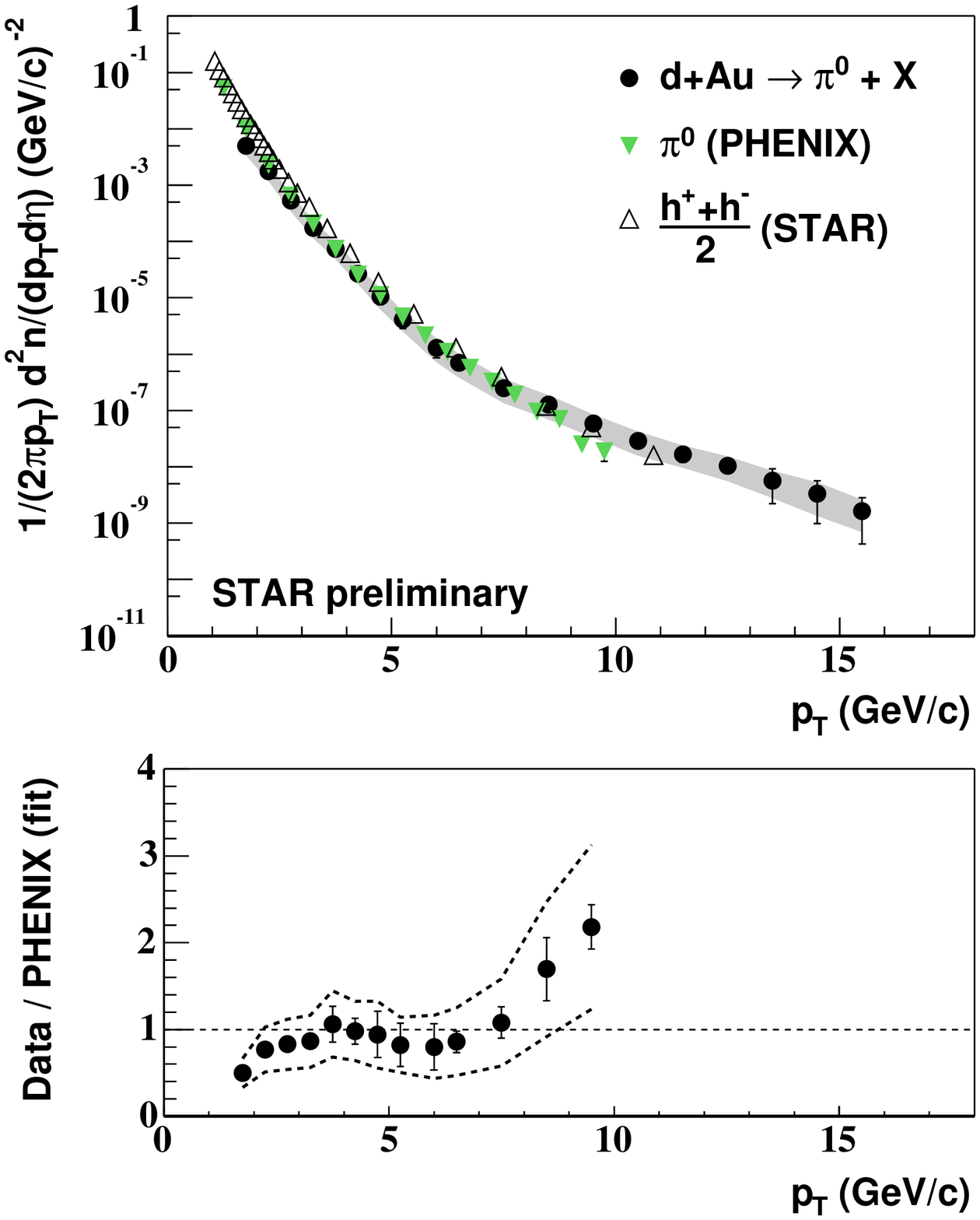}
 \end{center}
\end{minipage}
\hspace{2.3cm}
%
\begin{minipage}[b]{6cm}
 \begin{center}
 \includegraphics[height=8.8cm]{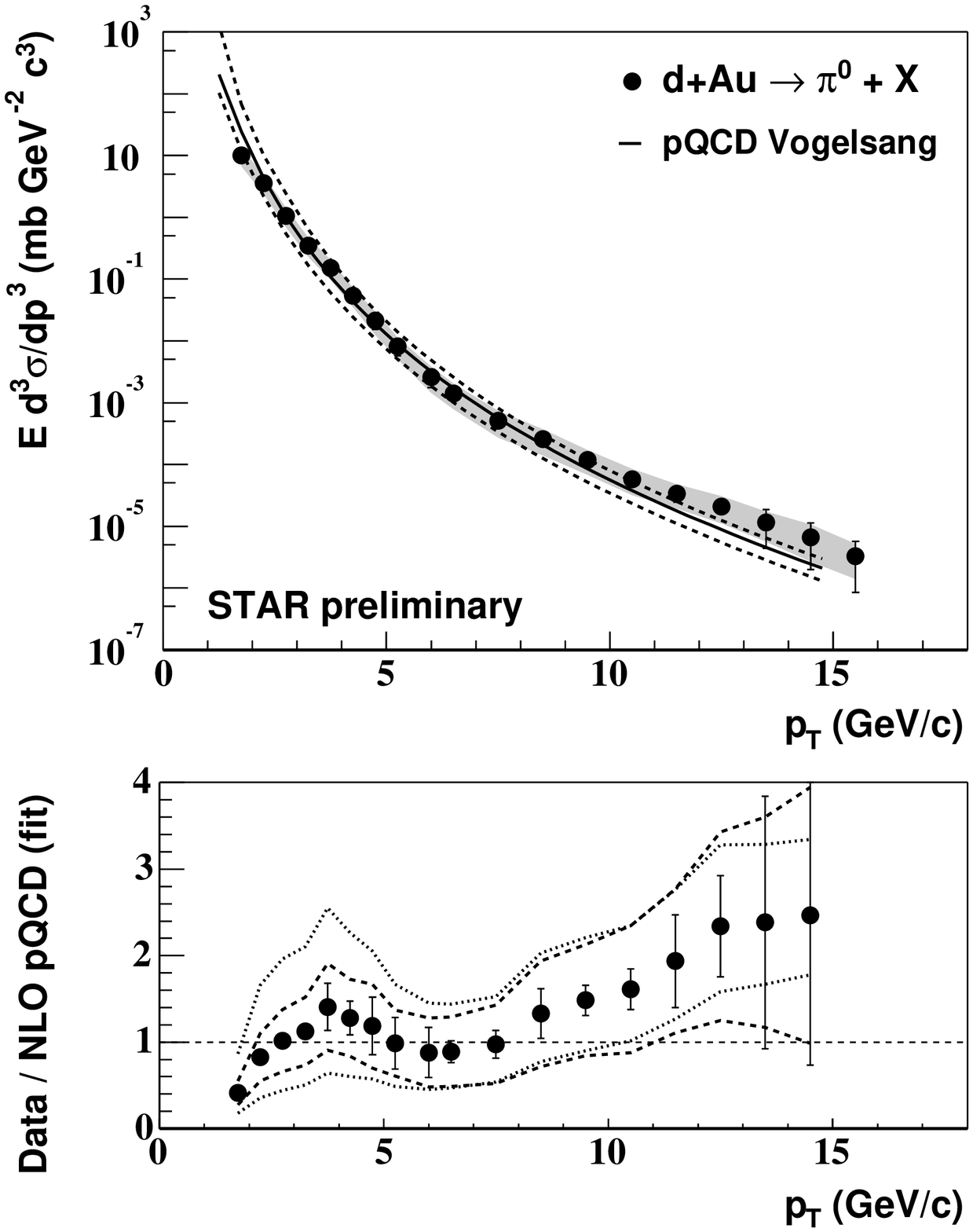}
 \end{center}
 \end{minipage}
 \vspace{-1.cm}
 \caption{\protect \footnotesize 
Left: Inclusive $p_T$ distribution for neutral pions in d-Au collisions at
$0<\eta<1$ (solid circles). 
The error bars (shaded bands) represent the statistical (systematic) 
uncertainties.
Previous STAR measurements on the charged hadron cross section~\cite{dAuStar03} 
are shown by the open triangles. 
The full triangles show $\pi^0$ results from PHENIX~\cite{dAuPhenix03}.
Right: The invariant differential cross section (full circles) compared to 
NLO pQCD calculations (solid line). 
The factorization scale uncertainty is illustrated by dashed line.
The lower plots show the ratios of the data to a fit of the PHENIX results 
(left) and to the theory curve (right). The dashed lines indicate 
the systematic errors derived from the data. The dotted curves (right) 
were obtained using different factorization scales.\vspace{-0.5cm}} 
 \label{fig2}
\end{figure}
In Fig.~\ref{fig1}, the resulting invariant mass distribution of cluster
pairs is shown. A clear signal is observed with a RMS width of 28 MeV/$c^2$.
The combinatorial background from random pairs was estimated by two different 
methods. First, a second-order polynomial was fitted to the invariant mass
distribution outside the peak region (solid curve in the left plot of 
Fig.~\ref{fig1}). 
The second method describes the background using the event mixing technique. 
The event mixing distribution, which is shown in the lower histogram in the 
left plot of Fig.~\ref{fig1}, describes the background reasonably well in the 
range 0.8--5 GeV/$c^2$ (not shown).
At lower invariant mass the excess can be attributed to the tail of the 
$\pi^0$ peak and a contribution from the $\eta$ signal 
($m_{\eta}$ = 547.3 MeV/$c^2$). The peak observed at $m <$ 0.05 GeV/$c^2$ 
stems from cluster splittings in the EMC towers.

The background subtracted spectra are shown in the middle and right-hand plot
of Fig.~\ref{fig1}. 
The yields per event obtained from both subtraction methods were extracted in 
$p_T$ bins (width of 0.5 GeV for minimum bias and 1 GeV for high tower triggered 
events) by integrating the background subtracted mass distribution in a range 
$\pm3\sigma$ around the $\pi^0$ peak. The mean values were used for further 
analysis and the difference contributes 10--15\% to the systematic uncertainties. 
The signal-to-background ratio increases from 1 at about $p_T$ = 1~GeV/c to 8 
at 4~GeV/c.

Corrections for reconstruction losses (quality cuts) and detector efficiencies 
were calculated with Monte-Carlo simulations using the STAR detector geometry 
and reconstruction software.
A correction for the unmeasured trigger fraction, which is expected to be a few 
percent, is not applied.
Losses due to cluster density effects and contributions from weak decays of K$^0$
mesons are not corrected yet, but are expected to be small.
The high tower trigger spectra are normalized using pre-scale factors obtained from 
the ratios of the BEMC cluster transverse energy distributions in the overlap region.
The overall systematic errors related to efficiency, yield extraction, pre-scale 
factors, and energy calibration are estimated to be 30\% for low and 50\% for high 
transverse momenta.

\section{Results and discussions}

The inclusive $p_T$ distribution for neutral pions is shown in the left panel
of Fig.~\ref{fig2}. 
The yields up to 6 GeV/c are from minimum bias events while above 6 (9.5)
GeV/c they are from HT1 (HT2) triggered events.
For the different trigger samples the yields have an overlap of one point 
in the $p_T$ spectrum and agree within errors.
It is seen from Fig.~\ref{fig2} that $\pi^0$ mesons are presently measured 
up to $p_T$ = 16 GeV/c.
The $p_T$ spectrum is compared with previous STAR measurements on charged 
hadron cross section~\cite{dAuStar03} and with PHENIX results on neutral 
pion production~\cite{dAuPhenix03}. 
There is a reasonable agreement within 10--20\%.

The right panel of Fig.~\ref{fig2} shows the invariant differential cross 
sections obtained from the product of the yields and the hadronic cross
section in d-Au collisions~\cite{dAuStar03}.
The normalization uncertainty is 10\%.
The results are compared to next-to-leading order (NLO) pQCD 
predictions~\cite{Vog04} which are calculated using the CTEQ6M set of nucleon 
parton distribution functions~\cite{CTEQ6M} and the nuclear parton densities 
in the gold nucleus from~\cite{AuPDF}.
In this calculation the factorization scale was identified with $p_T$ and
is varied by a factor two to estimate the scale uncertainties 
(see Fig.~\ref{fig2}). 
The fragmentation functions are taken from~\cite{KKP00}. The Cronin effect 
is not included in the calculations. The data, within errors, are consistent 
with the calculation up to $p_T$ = 15~GeV/c.

\begin{footnotesize}

\bibliography{../literatur_highpt.bib}

\begin{thebibliography}{10}

\bibitem{SuppPartSpez04}
J. Adams {\it et al.}~(STAR~Collaboration), Phys. Rev. Lett. {\bf 92},
  052302  (2004).

\bibitem{SuppPart03}
J. Adams {\it et al.}~(STAR~Collaboration), Phys. Rev. Lett. {\bf 91},
  172302  (2003).

\bibitem{BBCorr03}
C. Adler {\it et al.}~(STAR~Collaboration), Phys. Rev. Lett. {\bf 90},
  082302  (2003).

\bibitem{Flow03}
C. Adler {\it et al.}~(STAR~Collaboration), Phys. Rev. Lett. {\bf 90},
  032301  (2003).

\bibitem{Jac04}
P. Jacobs and X.~N. Wang, to be published in Prog. Part. and Nucl. Phys.
  (arXiv: hep-ph/0405125).

\bibitem{NimSTAR03}
K.~H. Ackermann {\it et al.}~(STAR~Collaboration), Nucl. Instrum. Meth.
  {\bf A499},  624  (2003).

\bibitem{NimEmc03}
M. Beddo {\it et al.}~(STAR~Collaboration), Nucl. Instrum. Meth. {\bf
  A499},  725  (2003).

\bibitem{NimAbsEnCal02}
T.~M. Cormier {\it et al.}~(STAR~Collaboration), Nucl. Instrum. Meth. {\bf
  A483},  734  (2002).

\bibitem{EtSTAR04}
J. Adams {\it et al.}~(STAR~Collaboration), arXiv: nucl-ex/0407003
  (submitted to Phys. Rev. C)  .

\bibitem{dAuStar03}
J. Adams {\it et al.}~(STAR~Collaboration), Phys. Rev. Lett. {\bf 91},
  072304  (2003).

\bibitem{dAuPhenix03}
S.~S. Adler {\it et al.}~(PHENIX~Collaboration), Phys. Rev. Lett. {\bf 91},
   072303  (2003).

\bibitem{Vog04}
W. Vogelsang, 2004, private communication.

\bibitem{CTEQ6M}
J. Pumplin {\it et al.}~(CTEQ~Collaboration), J. High Energy Phys. {\bf
  0207},  012  (2002).

\bibitem{AuPDF}
L. Frankfurt and M. Strikman, Eur. Phys. J. {\bf A5},  293  (1999), L.
  Frankfurt, V. Guzey, M. McDermott, and M. Strikman, J. High Energy Phys. {\bf
  0202}, 027 (2002), L. Frankfurt, V. Guzey and M. Strikman, arXiv:
  hep-ph/0303022.

\bibitem{KKP00}
B.~A. Kniehl, G. Kramer, and B. P\"otter, Nucl. Phys. {\bf B582},  514  (2000).

\end{thebibliography}


\end{footnotesize}

\end{document}